\documentstyle[multicol,pre,aps,epsf]{revtex}
\draft
\newcommand{\tr}{\mbox{Tr} }
\newcommand{\ket}[1]{\left | #1 \right \rangle}
\newcommand{\bra}[1]{\left \langle #1 \right |}
\newcommand{\proj}[1]{\ket{#1} \! \bra{#1}}

\begin{document}
\title{Noise effects on One-Pauli Channels}
\author{
Julian Juhi-Lian Ting
%\footnote{Electronic address: jlting@multimania.com}
}
\address{
jlting@multimania.com}
%No.38, Lane 93, Sec.2, Leou-Chuan E. Rd., Taichung, 40312 Taiwan, ROC}
\date{\today}
\maketitle
\begin{abstract}
The possibility of stochastic resonance of a quantum channel and
hence the noise enhanced quantum channel capacity is
explored by considering one-Pauli channels which are more classical like.
The fidelity of the channel is also considered.

\end{abstract}
\begin{multicols}{2}

\section{Introduction}

Recently because of the development of quantum computers\cite{DA}
people have  become interested
in information transmission through quantum channels\cite{S}. 
Quantum information theories\cite{CHB} can be used to describe processes 
such as data storage, quantum
cryptography\cite{DEJ} and quantum teleportation\cite{BBP}.
However, after an initial burst of papers following Shor's discover of
quantum factoring algorithm\cite{SH}, almost every work is aiming to solve
the decoherence problem which is responsible for transition into 
effectively classical behaviour\cite{JP}.
There are people using NMR techniques, which provide longer
decoherence time than previous techniques,
claiming they can built a quantum
computer with a cup of coffee\cite{GC}.
There are also people trying to use various software methods,
in particular, quantum error correcting codes, to correct decoherence
induced errors.
The decoherence problem lay at the heart of the development of
quantum mechanics.
Apparently, decoherence is a hurdle need to be surmounted 
before quantum computers can be 
materialized.
However, is decoherence, the counterpart of classical noise, really nuisance?
For people who know stochastic resonance\cite{BSV}, the answer is perhaps 'no'.
Decoherence can be perhaps used as a resource as entanglement had.

In two previous papers we have considered the noise effects on
the two-Pauli channels\cite{T} and the
depolarizing channels\cite{TT} using the concept from stochastic resonance.
%We found the asymmetry of the channel is important for some interesting
%effects.
There is a question that whether does the results obtained so far consistent
with the classical results?
For the two-Pauli channel and depolarizing channel, it is not possible to
check because they consist of flipping the phase of the qubits
(quantum bits) \cite{BS}
which has no classical counter part.
However, it might be possible
to consider a one-Pauli channel that only flip the qubits amplitude. 
The $\sigma_1$ channel can be viewed as a binary symmetric channel,
while $\sigma_2$ and $\sigma_3$ channels are $\sigma_1$ channel
in a dual basis.
Furthermore, we still do not know what are the noise effects for each 
Pauli operator separately. 
Therefore, in this paper we will consider three different
one-Pauli channels to see how will the noise influence their
capacity and fidelity.

\section{the Noisy Channel Model}
The classical world is made from different material. 
However, in the quantum world all objects are 
made of the same elementary particles.
The particles are in different states of superposition.
Only the information describing them are different.

In the classical world the 
information is coded as {\it bits} and is described by $0$ or $1$, 
while the quantum world the information is coded as {\it  qubit}
and is described by the corresponding {\it density matrix}.

Schumacher and Nielsen \cite{S96b} have developed a quantum information
theory to describe the information processing in the
quantum world. In their formulation
a quantum channel  can be considered as
a process defined by an input density matrix 
$\rho_x$, and an output density matrix $\rho_y$, with the process
described by a quantum operation, ${\cal N}$,
\begin{eqnarray} 
\rho_x \stackrel{{\cal N}}{\rightarrow} \rho_y.
\end{eqnarray}
Because of decoherence, the super-operator $\cal N$
is not unitary.
However,
on a larger quantum system that includes
the environment $E$ of the system, the total evolution
operator $U_{x E}$ become unitary.  This environment may be considered
to be initially in a pure state $\ket{0_{E}}$ without
loss of generality.  In this case, the
super-operator can be written as
\begin{equation}
        {\cal N} (\rho_{x}) = \tr_{E} U_{xE} \left (
                                  \rho_{x} \otimes \proj{0_{E}} \right )
                                  {U_{xE}}^{\dagger} .
\label{channel}
\end{equation}
The partial trace, $\tr_E$, is taken over environmental degree of freedom,
and $\otimes$ means a direct product for the spaces on both sides of the
operator.
Eq.~(\ref{channel}) can be rewritten as a completely positive linear
transformation acting on the density matrix:
\begin{equation}
{\cal N} (\rho_x)=\sum_i A_i\rho_x A_i^\dagger\;,
\label{AnklesOfHair}
\end{equation}
in which the $A_i$ satisfy the completeness relation
\begin{equation}
\sum_i A_i^\dagger A_i = I\;,
\label{HankThoreau}
\end{equation}
which is equivalent to requiring $\tr [{\cal N} (\rho_x )]=1$.
The mutual information of a classical channel with
classical sources,  written using quantum formalism become\cite{BNS}
\begin{eqnarray}
H(x:y) = H(\rho_x)+H({\cal N}(\rho_x))-H_e(\rho_x,{\cal N}),
\end{eqnarray}
in which
%$H ( \cdot ) = - \tr \cdot \log_2 \cdot $ is the von Neumann entropy\cite{N},
$H ( \bullet ) = - \tr \left[ \bullet \log_2 \bullet \right]$ 
is the von Neumann entropy\cite{N},
and 
\begin{eqnarray}
H_e(\rho_x,{\cal N}) \equiv - \tr (W \log_2 W), \end{eqnarray}
with 
\begin{eqnarray} 
W_{ij} \equiv \mbox{Tr}(A_i \rho_x A_j^{\dagger})
\end{eqnarray}
measures the amount of information 
exchanged between the system $x$ and the environment $E$ during
their interaction\cite{S}, which 
can be used to characterize the amount of quantum noise, $N$, in the channel.
If the environment is initially in a pure state,
the entropy exchange is just the environment's entropy after the
interaction.
Hence, the coherent information is defined as
\begin{eqnarray}
C (\rho_x,{\cal N}) \equiv H \left(
        {\cal N}(\rho_x) \right) -
        N (\rho_x,{\cal N}),
\end{eqnarray}
which plays a role in quantum information theory analogous to that played
by the mutual information in classical information theory.

\section{One-Pauli Channels}
In what follows the influence of noise on three kinds of
one-Pauli channels with a general input state 
\begin{equation}
\rho_x=\frac{1}{2}\Big(I + \vec{a}\cdot\vec{\sigma}\Big)\;.
\end{equation}
is considered.
Here, $I$ is the identity matrix,
$\vec{a}=(a_1,a_2,a_3)$ is the Bloch vector of length unity or less,
and $\vec{\sigma}$ is the vector of Pauli matrices,
which are defined as
\begin{equation}
\sigma_1=\left(\begin{array}{cr}
0 & 1\\
1 & 0\end{array}\right),\;\;\;\;
\sigma_2=\left(\begin{array}{cr}
0 & -i\\
i & 0\end{array}\right),\;\;\;\;
\sigma_3=\left(\begin{array}{cr}
1 & 0\\
0 & -1\end{array}\right).
\end{equation}
The output of the channel can always be written as
\begin{equation}
{\cal N}(\rho_x)=
\frac{1}{2}\Big(I + \vec{b}\cdot\vec{\sigma}\Big)\;.
\end{equation}
Two more definitions are needed in our computing for the channel properties:
The von Neumann entropy is
\begin{equation}
H (\rho_x) = - \sum_i \theta_i \log \theta_i,
\end{equation}
in which $\theta_i$s are the eigenvalues of the density matrix $\rho_x$.
Furthermore,
the (entangled) fidelity
\begin{equation}
F = \sum_{\mu} (\tr \rho_x A_{\mu})(\tr \rho_x A_{\mu}^{\dagger}),
\end{equation}
is also of our concern, since it represent the quality of the signal 
transmitted. 

\subsection{$\sigma_1$ channel}
\begin{figure}
\epsfxsize=7.5cm\epsfbox{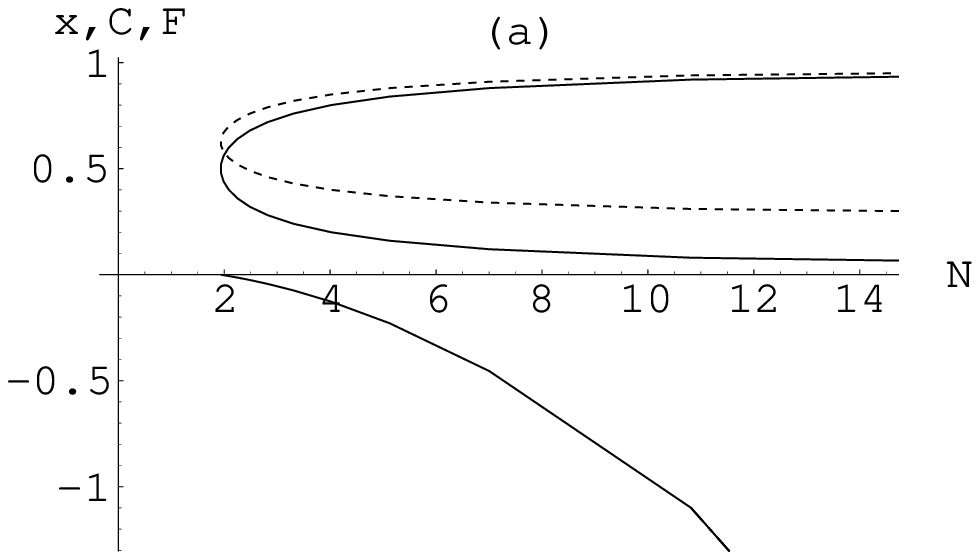}
\epsfxsize=7.5cm\epsfbox{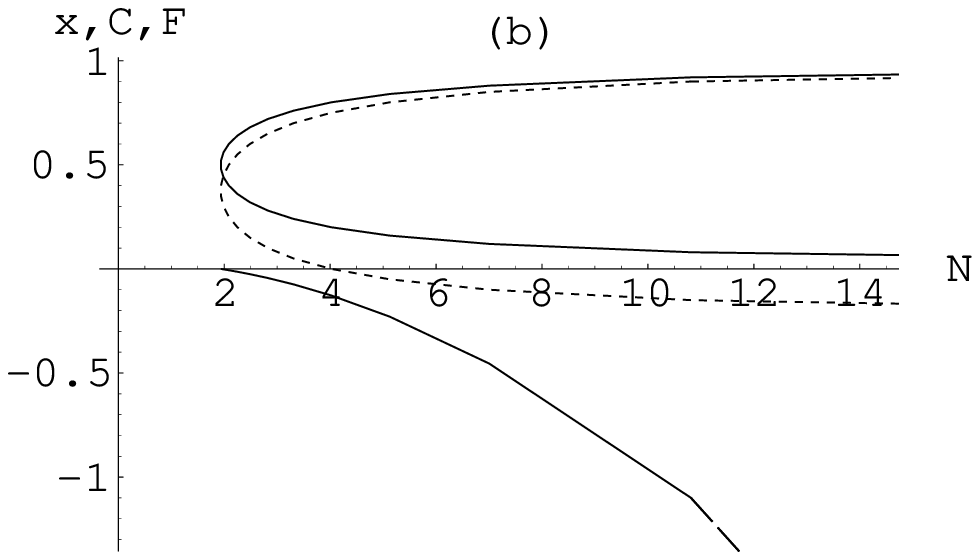}
\epsfxsize=7.5cm\epsfbox{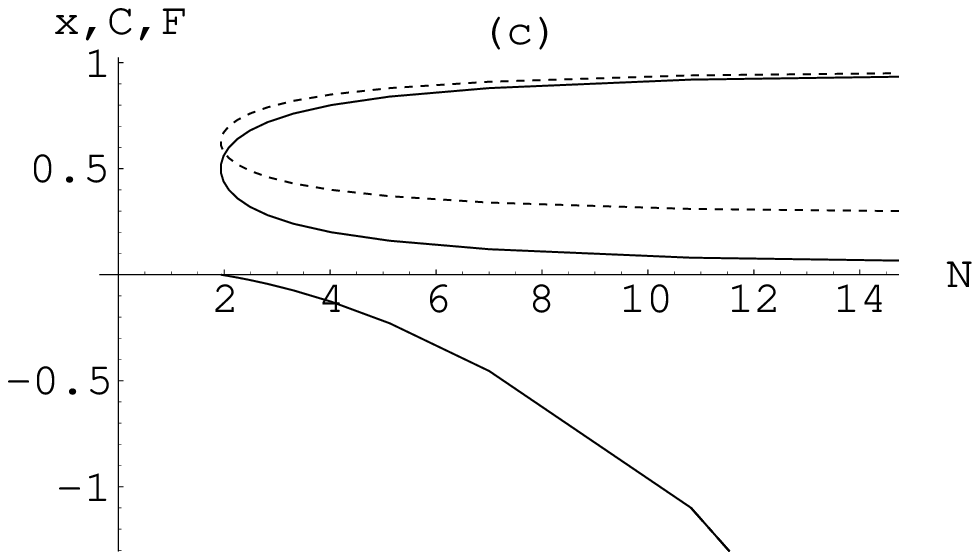}
\vskip 0.3cm
\caption{Parametric plots of the 
retention rate, $x$, versus noise, $N$ (solid lines);
coherent information, C, versus
noise, N, (long dashed lines)
and fidelity, F, versus noise, N, (short dashed lines) for the parameter
$x$ from $0$ to $1$ and
(a) $\sigma_1$ channel with initial state $a_1 = 5/10, a_2 = 6/10, a_3 =6/10;$
(b) $\sigma_2$ channel with initial state $a_1 = 6/10, a_2 = 5/10, a_3 =6/10;$
(c) $\sigma_3$ channel with initial state $a_1 = 6/10, a_2 = 6/10, a_3 =5/10;$
}
\label{fig1}
\end{figure}

A $\sigma_1$ channel
can be written in terms of $A_i$'s in Eq.~(\ref{AnklesOfHair}) as
\begin{equation}
A_1=\sqrt{x}\,I\;,\;\;\;\;
A_2=\sqrt{1-x}\,\sigma_1\;,\;\;\;\;
\end{equation}
This channel 
flips the qubit amplitude with probability $1-x$.
We have
\begin{equation}
\vec{b}=\Big(a_1,\,a_2 (2 x -1),\,a_3(2 x -1)\Big)\;.
\end{equation}
The matrix $W$ for the $\sigma_1$ channel read,
\begin{equation}
W=\left(\begin{array}{cc}
x & a_1 \sqrt{x (1-x)}\\
a_1 \sqrt{x (1-x)}& {1-x}
\end{array}\right).
\end{equation}
It eigenvalues are
\begin{equation}
\lambda_{1,2}= \left[1 \pm \sqrt{1-4 x (x-1)(a_1^2 - 1)} \right]/2.
\end{equation}
Hence,
\begin{equation}
N  = - \sum_{i=1}^2 \lambda_i \log_2 \lambda_i,
\end{equation}
while
$\theta_{1,2} = \left[ 1\pm \sqrt{a_1^2+(a_2^2+a_3^2)(1-2x)^2} \right]/2$.
This $x-N$, $C-N$ relationship is plotted in Fig.~\ref{fig1}(a).

For the $\sigma_1$ channel the entangled fidelity
\begin{equation}
F = a_1^2  (1 - x)  + x.
\end{equation}
The relation between the
fidelity and the noise is  plotted with the coherent information
in Fig.~\ref{fig1}(a). 

\subsection{$\sigma_2$ channel}

Similarly, a $\sigma_2$ channel
can be written in terms of $A_i$'s in Eq.~(\ref{AnklesOfHair}) as
\begin{equation}
A_1=\sqrt{x}\,I\;,\;\;\;\;
A_2=-i \sqrt{ 1-x}\,\sigma_2\;,\;\;\;\;
\end{equation}
This channel 
flips the qubit amplitude and phase with probability $1-x$.
The action of the channel on this density matrix is:
\begin{equation}
{\cal N}(\rho_x)=
\frac{1}{2}\Big(I + \vec{b}\cdot\vec{\sigma}\Big)\;,
\end{equation}
in which 
\begin{equation}
\vec{b}=\Big(a_1 (2x-1),\,a_2,\,a_3 (2x-1) \Big)\;.
\end{equation}
The matrix $W$ for the $\sigma_2$ channel read,
\begin{equation}
W=\left(\begin{array}{cc}
x & i a_2 \sqrt{x (1-x)}\\
-i a_2 \sqrt{x (1-x)}& {1-x}
\end{array}\right).
\end{equation}
It eigenvalues are
\begin{equation}
\lambda_{1,2}= \left[1 \pm \sqrt{1-4 x (x-1)(a_2^2 - 1)} \right]/2.
\end{equation}
Hence,
\begin{equation}
N  = - \sum_{i=1}^2 \lambda_i \log_2 \lambda_i,
\end{equation}
while
$\theta_{1,2} = \left[ 1\pm \sqrt{a_2^2+(a_1^2+a_3^2)(1-2x)^2} \right]/2$.
For the $\sigma_2$ channel
\begin{equation}
F = - a_2^2  (1 - x)  + x.
\end{equation}
The relation between the
fidelity and the noise is  plotted with the coherent information
in Fig.~\ref{fig1}(b). 

\subsection{$\sigma_3$ channel}
A $\sigma_3$ channel
can be written as
\begin{equation}
A_1=\sqrt{x}\,I\;,\;\;\;\;
A_2=\sqrt{ 1-x}\,\sigma_3\;,\;\;\;\;
\end{equation}
This channel 
flips the qubit phase with probability $1-x$.
The action of the channel on this density matrix is:
\begin{equation}
{\cal N}(\rho_x)=
\frac{1}{2}\Big(I + \vec{b}\cdot\vec{\sigma}\Big)\;,
\end{equation}
in which 
\begin{equation}
\vec{b}=\Big(a_1 (2x-1),\,a_2 (2 x -1),\,a_3\Big)\;.
\end{equation}
The matrix $W$ for the $\sigma_3$ channel read,
\begin{equation}
W=\left(\begin{array}{cc}
x & a_3 \sqrt{x (1-x)}\\
a_3 \sqrt{x (1-x)}& {1-x}
\end{array}\right).
\end{equation}
It eigenvalues are
\begin{equation}
\lambda_{1,2}= \left[1 \pm \sqrt{1-4 x (x-1)(a_3^2 - 1)} \right]/2.
\end{equation}
Hence,
\begin{equation}
N  = - \sum_{i=1}^2 \lambda_i \log_2 \lambda_i,
\end{equation}
while
$\theta_{1,2} = \left[ 1\pm \sqrt{a_3^2+(a_1^2+a_2^2)(1-2x)^2} \right]/2$.
For the $\sigma_3$ channel
\begin{equation}
F = a_3^2  (1 - x)  + x.
\end{equation}
The relation between the
fidelity and the noise is  plotted with the coherent information
in Fig.~\ref{fig1}(c). 

We noticed that for all three channels
the capacity is always  non-positive whatever the amount of
noise is. 
\section{conclusion}
In conclusion we found, as far as the coherent information
and fidelity are concerned,
these three one-Pauli channels are the same except for the fidelity
of the $\sigma_2$ channel.
This is perhaps the reason that we found in the previous works
why there is noise enhancement of the fidelity but
could not find  noise enhancement for the coherent information.

%%%%%%%%%%%%%%%%%%%%%%%%%%%%%%%%%%%%%%%%%%%%%%%%%%%%%%%%%%%%%%%%%%%%%%%%%%%%%%%
\end{multicols}
%%%%%%%%%%%%%%%%%%%%%%%%%%%%%%%%%%%%%%%%%%%%%%%%%%%%%%%%%%%%%%%%%%%%%%%%%%%%%%%
\end{document}